\documentclass[aps,prd,amsmath,twocolumn,showpacs]{revtex4}

\usepackage{epsfig}
\usepackage{graphics}
\usepackage{latexsym}
\usepackage{amsmath}
\usepackage{amssymb}
\usepackage{rotating}
\usepackage{subfigure}
\usepackage{bm}
\usepackage{color}

\usepackage[colorlinks = true,
linkcolor = magenta,
urlcolor  = blue,
citecolor = red,
anchorcolor = blue]{hyperref}

\begin{document}

\title{  Improved determination of $\bar d(x) - \bar u(x)$ flavor asymmetry in the proton by BONuS experiment at JLAB and using an approach by Brodsky, Hoyer, Peterson, and Sakai }

\author{Maral Salajegheh$^{a}$}
\email{M.Salajegheh@stu.yazd.ac.ir}

\author{Hamzeh Khanpour$^{b,c}$}
\email{Hamzeh.Khanpour@mail.ipm.ir}

\author{S. Mohammad Moosavi Nejad$^{a,c}$}
\email{Mmoosavi@yazd.ac.ir}

\affiliation {
$^{a}$Physics Department, Yazd University, P.O.Box 89195-741, Yazd, Iran   \\
$^{b}$Department of Physics, University of Science and Technology of Mazandaran, P.O.Box 48518-78195, Behshahr, Iran  \\
$^{c}$School of Particles and Accelerators, Institute for Research in Fundamental Sciences (IPM), P.O.Box 19395-5531, Tehran, Iran 
 }

\date{\today}

%
\begin{abstract}

The experimental data taken from both Drell-Yan and deep-inelastic scattering (DIS) experiments 
suggest a sign-change in $\bar d(x) - \bar u(x)$ flavor asymmetry in the proton at large values of momentum fraction $x$. 
In this work, we present a phenomenological study of $\bar d(x) - \bar u(x)$ flavor asymmetry.
First, we extract the $\bar d(x)-\bar u(x)$ distribution using the more recent data from the BONuS experiment at Jefferson Lab on the ratio of neutron to proton structure functions, $F_2^n/F_2^p$,
and show that it undergoes a sing-change and becomes negative at large values of momentum fraction $x$, as expected. 
The stability and reliability of our obtained results have been examined by including target mass corrections (TMCs) as well as higher twist (HT) terms which are particularly important
at the large-$x$ region at low Q$^2$.
Then, we calculate the $\bar d(x) - \bar u(x)$ distribution 
using the Brodsky, Hoyer, Peterson, and Sakai (BHPS) model and show that if one chooses a mass for the down quark smaller than the one for the up quark it leads to a better description for the Fermilab E866 data. In order to prove this claim, we determine the masses of down and up sea quarks by fitting to the available and up-to-date experimental data for the $\bar d(x)-\bar u(x)$ distribution. In this respect, unlike the previous performed theoretical studies, we have shown that this distribution has a sign-change at $x>0.3$ after evolution to the scale of available experimental data. 

\end{abstract}


\maketitle

\tableofcontents{}

%
\section{Introduction}\label{sec:one}

The parton distribution functions (PDFs) content for nucleon is usually determined from global fits to experimental data at the large momentum transfer Q$^2$. 
Over the past decade, our knowledge of the quark and gluon substructure of the nucleon has been extensively improved due to the high-energy scattering data from the fixed target experiments, the data from $ep$ collider HERA~\cite{Abt:2016zth,Abramowicz:2015mha,South:2016cmx} and also from high energy $p \bar p$ scattering at the Tevatron~\cite{Abazov:2008ae,Aaltonen:2008eq}.
More recently, the data taken from various channels in $pp$ collisions at the CERN LHC play a main role to constrain the sea quarks and gluon  distributions at the proton~\cite{Rojo:2015acz}. 
In recent years, various up-to-date efforts are being made to extract more complete information about the nucleon's quark and gluon structure in the form of parton distribution functions
for the unpolarized PDFs~\cite{Alekhin:2017kpj,Ball:2014uwa,Harland-Lang:2014zoa,Jimenez-Delgado:2014twa,Dulat:2015mca,Khanpour:2016uxh,MoosaviNejad:2016ebo,Goharipour:2017rjl} and the polarized PDFs~\cite{Jimenez-Delgado:2014xza,Sato:2016tuz,Shahri:2016uzl,Khanpour:2017cha,Ethier:2017zbq,Khanpour:2017fey,AtashbarTehrani:2013qea} cases.
These analysis are mainly focused on the extraction of the parton distribution functions at small and large values of $x$ up to next-to-next-to-leading order (NNLO) accuracy. Similar efforts have also been performed for the case of fragmentation functions (FFs)~\cite{Nejad:2015fdh,deFlorian:2017lwf,MoosaviNejad:2016qdx,Ethier:2017zbq,Bertone:2017tyb,Zarei:2015jvh,Boroun:2015aya,Boroun:2016zql}, nuclear PDFs~\cite{Klasen:2017kwb,Eskola:2016oht,Khanpour:2016pph,Kovarik:2015cma,Goharipour:2017uic} and generalized parton distributions (GPDs)~\cite{Kumericki:2016ehc,Khanpour:2017slc,Kumericki:2009uq,Mueller:2013caa}.

Since the Gottfried sum rule~\cite{Gottfried:1967kk} has been proposed in 1967, many experimental and theoretical researches have been widely performed
so far to check the validity or violation of it and also to study the antiquark flavor asymmetry $ \bar d-\bar u $ in the nucleon sea (see Ref.~\cite{Chang:2014jba} and references therein). 
If we adopt that the $\bar u$ and $\bar d$ distributions in the nucleon are the same and the isospin invariance is also valid, then the Gottfried sum rule is obtained 
by integrating the difference between the $F_2$ structure functions of the proton and neutron over $ x $ as 
$I_G\equiv \int^1_0 [F^p_2(x)-F^n_2(x)]/x~dx = 1/3$, where $ x $ is the Bjorken scaling variable. However, assuming the flavor
asymmetry of the nucleon sea, the Gottfried sum rule is violated by an extra term as $2/3\int^1_0 [\bar u(x) - \bar d(x)] dx$. In this way,
if there is a $ \bar d $ excess over $ \bar u $ in the nucleon, we expect a smaller value for the Gottfried sum than $ 1/3 $.

In 1991, the New Muon Collaboration (NMC) obtained the value $I_G = 0.235 \pm 0.026$ in measuring the proton and deuteron 
$ F_2 $ structure functions~\cite{Amaudruz:1991at} from deep-inelastic muon scattering on hydrogen and deuterium targets, which is approximately 
28\% smaller than the Gottfried sum. This measurement provided the first clear evidence for the breaking of this sum rule. In addition to the 
deep-inelastic scattering (DIS) experiments, the violation of the Gottfried sum rule can be investigated from 
semi-inclusive DIS (SIDIS) and Drell-Yan cross section measurements.
The related study has been performed by the HERMES collaboration~\cite{Ackerstaff:1998sr} in the case of SIDIS experiment. In this study a measurement of $\bar d(x) - \bar u(x)$ was reported over the range of $0.02 < x < 0.3$,
but with a rather large experimental uncertainty. On the other hand, the NA51~\cite{Baldit:1994jk} and FNAL E866/NuSea~\cite{Hawker:1998ty} collaborations studied this violation
by measuring $pp$ and $pd$ Drell-Yan processes and established again that there is a $\bar d$ excess over $\bar u$ in the nucleon sea. 
Although, the ratio of $\bar d / \bar u$ was only measured at the mean $x$-value of $\langle x\rangle $=0.18 in the NA51 experiment. The $x$-dependence of this ratio and the $\bar d(x) - \bar u(x)$ flavor asymmetry have also been measured over the kinematic region $0.015 < x < 0.35$ in the Fermilab E866 experiment.

In addition to the violation of the Gottfried sum rule as well as the existence of the $\bar d-\bar u$ flavor asymmetry in the nucleon sea,
one could take another important result from the Fermilab E866 data. In fact, the last data point suggested a sign-change for the $\bar d(x) - \bar u(x)$ distribution at $x \sim 0.3$, despite of their large uncertainty. To be more precise, it indicates that this distribution must be negative at the $x$-values
approximately larger than $0.3$. This can be very important issue because the perturbative regime of quantum chromodynamics (QCD) can not lead 
to a remarkable flavor asymmetry in the nucleon sea. Furthermore, according to the studies which have yet been  performed (for a review see Refs.~\cite{Chang:2014jba,Kumano:1997cy,Garvey:2001yq,Peng:2014hta}) the current theoretical models,
regardless of their ability to describe an enhancement of $\bar d$ over $\bar u$, can not predict a negative 
value for the $\bar d(x) - \bar u(x)$ distribution at any value of $x$. These theoretical studies are based on the various models such as Pauli-blocking~\cite{Field:1976ve,Schreiber:1991tc,Buccella:1992zs,Steffens:1996bc}, meson-cloud~\cite{Thomas:1983fh,Speth:1996pz,Alwall:2005xd,Traini:2013zqa},
chiral-quark~\cite{Szczurek:1996tp,Song:2011fc,Salamu:2014pka}, chiral-quark soliton~\cite{Pobylitsa:1998tk,Wakamatsu:2003wg,Wakamatsu:2009fn,Wakamatsu:2014asa},
intrinsic sea~\cite{Chang:2011vx,Chang:2011du,Chang:2014lea} and statistical~\cite{Bourrely:1994nm,Bourrely:2005kw,Zhang:2008nr} 
models. Except the Pauli-blocking model which considers a perturbative mechanism to describe the enhancement of $\bar d$ over $\bar u$, other models consider a nonperturbative origin for this effect and are almost successful. However, the Pauli-blocking model is not successful to produce the distribution of the $\bar d(x) - \bar u(x)$ when it is compared with the experimental data.

Recently, Peng \textit{et al.}~\cite{Peng:2014uea} have presented an independent evidence for the $\bar d(x) - \bar u(x)$ sign-change at $x \sim 0.3$ by analyzing the DIS data. 
They have showed that in addition to the Drell-Yan data, the analysis of the NMC DIS data for the $F^p_2 - F^n_2$~\cite{Amaudruz:1991at} and $F_2^d/F_2^p$~\cite{Arneodo:1996qe} can also lead to a negative value for the 
$\bar d(x) - \bar u(x)$ at $x \gtrsim 0.3$. They have also discussed the significance of this sign-change and the fact that none of the current theoretical models
can predict this asymmetry. 
Future Drell-Yan experiment at J-PARC P04~\cite{jpar} and also Fermilab E906~\cite{fermi} experiments will give us more accurate information on the $\bar d-\bar u$ flavor asymmetry, especially, at the larger values of $x$. This motivates us to study on this topic. 

In the present paper, following the studies performed by Peng \textit{et al.} for the extraction of $\bar d(x) - \bar u(x)$, we first investigate whether such behavior can be seen in the analysis of data from other experiments. If it is, we shall study the approximate position of the $\bar d(x) - \bar u(x)$ sign-change in $x$ and also estimate the magnitude of its negative area. 
In addition, since our study is in the low Q$^2$ at high value of $x$ in which the target mass corrections (TMCs) and higher twist (HT) effects are significant, then we develop our
analysis by considering these nonperturbative contributions.
Therefore, we calculate the $\bar d(x) - \bar u(x)$ distribution using the Brodsky, Hoyer, Peterson, 
and Sakai (BHPS) model~\cite{Brodsky:1980pb} and show that the available experimental data for this quantity suggest a smaller value for the down quark mass than the up quark one in the BHPS formalism. Note that, this is in contrast to the previous studies in this context~\cite{Chang:2011vx,Chang:2011du,Chang:2014lea} where was  assumed 
equal masses for the down and up quarks in the proton. This difference between masses leads to a sign-change for $\bar d(x) - \bar u(x)$ when we evolve this quantity to the scale of experimental data~\cite{Hawker:1998ty}.

The content of the present paper goes as follows: 
we compare the Fermilab E866~\cite{Hawker:1998ty} data with the prediction of the latest parton distribution functions from various groups 
and also extract the $\bar d(x) - \bar u(x)$ using the updated CLAS collaboration data for the $ F_2^n/F_2^p $ ratio in Sec.~\ref{sec:two}.
This section also includes detailed discussions on the nuclear corrections as well as the effects arising from the nonperturbative TMCs and HT terms.
In Sec.~\ref{sec:three}, we briefly introduce the BHPS model and explain the idea for choosing a smaller mass for the down quark than up quark in the BHPS formalism.
Then, we prove our claim and determine the masses of down and up sea quarks by fitting the available experimental data for the $\bar d(x) - \bar u(x)$.
Finally, we summarize our results and present our conclusions in Sec.~\ref{sec:four}. Appendix presents our {\tt FORTRAN} package containing the $\bar d$ and $\bar u$ intrinsic distributions using the BHPS model.

%
\section{$\bar d(x) - \bar u(x)$ from recent CLAS data}\label{sec:two}

\begin{figure}[t!]
\centering
\vspace{0.5 cm}
\includegraphics[width=8.0cm]{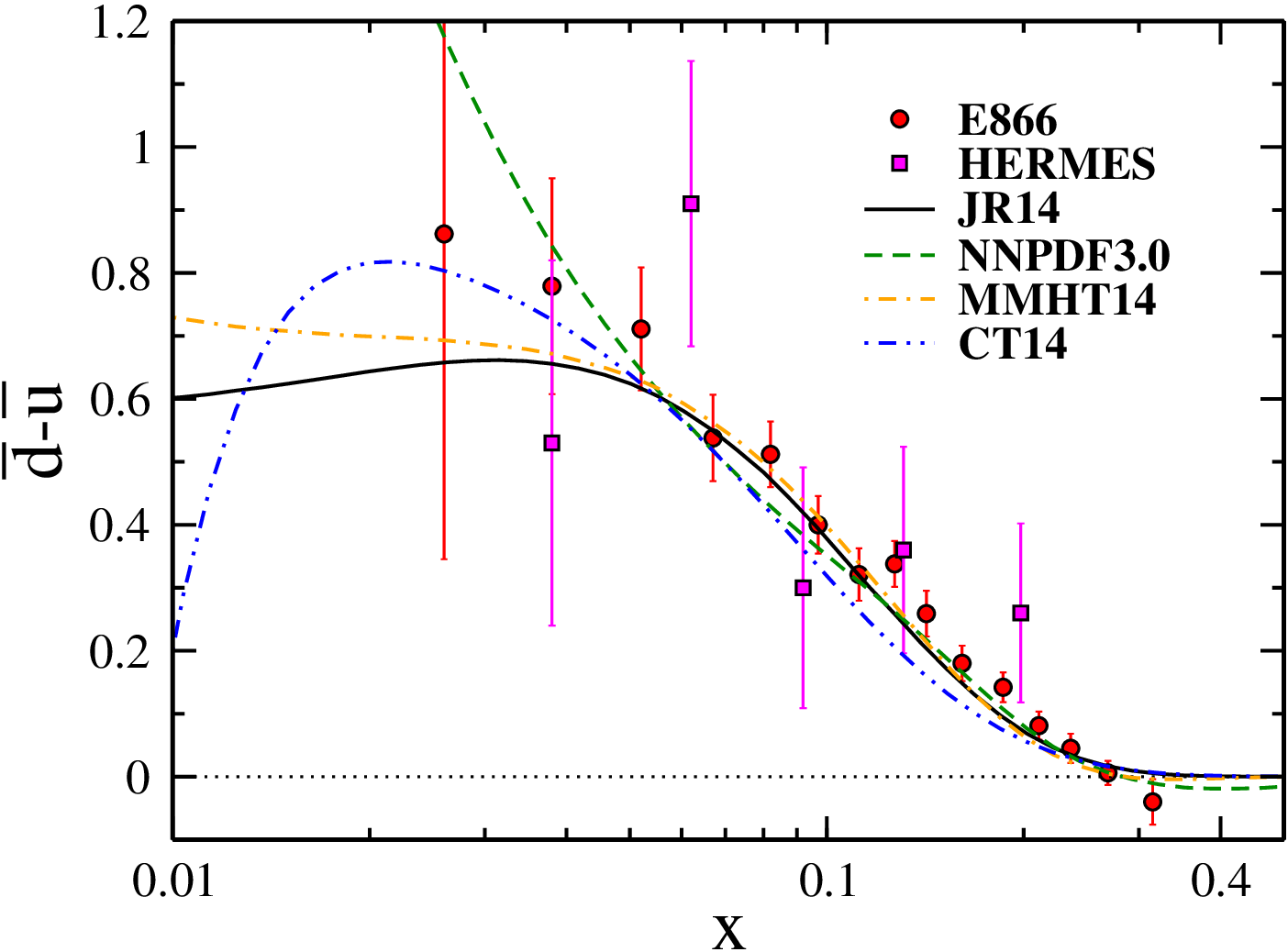}
\caption{  A comparison between HERMES~\cite{Ackerstaff:1998sr} and Fermilab E866~\cite{Hawker:1998ty} collaborations data for the $\bar d(x) - \bar u(x)$
and the NNLO theoretical predictions of JR14~\cite{Jimenez-Delgado:2014twa}, NNPDF3.0~\cite{Ball:2014uwa}, MMHT14~\cite{Harland-Lang:2014zoa} and CT14~\cite{Dulat:2015mca}
PDFs at Q$^2 = 54 \, {\rm GeV}^2$.  }
\label{fig:fig1}
\end{figure}

In recent years, our knowledge of the nucleon structure have been developed to a large extent, but it is not still enough. In this respect, an updated global analysis of PDFs including
a broad range of the experimental data from the various observables and also theoretical improvements can play an important role. In the theoretical studies, generally, an independent
parametrization form is chosen for the $\bar d(x) - \bar u(x)$ distribution in the global analysis of PDFs at the initial scale $Q_0$.
Fig.~\ref{fig:fig1} shows the $\bar d(x) - \bar u(x)$ data from the HERMES and the Fermilab E866 at Q$^2 = 2.5$ and $54$ GeV$^2$, respectively, which have been 
compared with the NNLO theoretical predictions of JR14~\cite{Jimenez-Delgado:2014twa}, NNPDF3.0~\cite{Ball:2014uwa}, MMHT14~\cite{Harland-Lang:2014zoa} and CT14~\cite{Dulat:2015mca}
PDFs for Q$^2 = 54$ GeV$^2$. Although, all predictions are in good agreement with these data, but they have major differences
with each other. For example, there is no possibility to change the $\bar d(x) - \bar u(x)$ sign at large-$x$ in JR14 parametrization 
unlike other PDF sets or the CT14 parametrization predicts $\bar d(x) - \bar u(x) < 0$ in small $x$-region. There is also another important conclusion which can be
taken from the E866 data. As it is clear from Fig.~\ref{fig:fig1} the last data point, despite of its large uncertainty, indicates that the $\bar d(x) - \bar u(x)$ must be negative at $x$-values
approximately larger than $0.3$.

Recently, Peng \textit{et al.}~\cite{Peng:2014uea} showed that in addition to the Drell-Yan data, there is an independent evidence for the $\bar d(x) - \bar u(x)$ sign-change at $x \sim 0.3$. Their results have been achieved by analyzing the NMC DIS data for the $F^p_2 - F^n_2$~\cite{Amaudruz:1991at} and the $F_2^d/F_2^p$~\cite{Arneodo:1996qe}. In this section, we are going to investigate  if such behavior can be seen in the analysis of data from other experiments such as Barely Off-shell Nucleon Structure (BONuS) experiment at Jefferson Lab. In this way, we can compute the position of the $\bar d(x) - \bar u(x)$ sign-change in $x$ and it is also possible to estimate the magnitude of its negative area.

From the parton model, one knows that the $F_2^{p,n}$ structure function of the nucleon at the leading-order (LO) of strong coupling constant $\alpha_s$ is expressed as an expansion of parton distributions $f_i(x)$, $F_2^{p,n}(x) = \sum_i e_i^2 \, xf_i(x) $, where $i$ denotes the flavor of the quarks and $e_i$ is the charge of $i$'th quark. It should be noted
that, in general, the parton distributions and in conclusion the structure functions depend on the four-momentum transfer squared $Q^2$. Now, if we adopt the charge symmetry 
of parton distributions in proton and neutron and also assume that the perturbatively generated $s, c, b$ quark distributions are equal in different
nucleons, the following relation is obtained for the $F^p_2 - F^n_2$ at LO

\begin{equation}
F^p_2(x) - F^n_2(x) = \frac{1}{3} x[u(x) + \bar u(x) - d(x) - \bar d(x)].
\label{eq1}
\end{equation}
In consequence, using the definition of valence quark, $q_v = q - \bar q$, the above relation  can be used to extract the $\bar d(x) - \bar u(x)$ 
as follows
\begin{equation}
\bar d(x) - \bar u(x) = \frac{1}{2} [u_v(x) - d_v(x)] - \frac{3}{2x}[F^p_2(x) - F^n_2(x)].
\label{eq2}
\end{equation}

According to Eq.~\eqref{eq2}, having two quantities $u_v(x) - d_v(x)$ and $F^p_2(x) - F^n_2(x)$ for a given value of $x$, one can extract the $\bar d(x) - \bar u(x)$ flavor asymmetry. 
For the first term in Eq.~(\ref{eq2}), we can use the related parameterizations from the various PDFs~\cite{Jimenez-Delgado:2014twa,Ball:2014uwa,Harland-Lang:2014zoa,Dulat:2015mca}
and the last term ($F^p_2(x) - F^n_2(x)$) in the second bracket can be calculated, for example, 
from the new CLAS Collaboration data reported for the $F_2^n/F_2^p$~\cite{Tkachenko:2014byy}. Since we are looking for a possible
sign-change in the $\bar d(x) - \bar u(x)$ at a large value of $x$, in this work we use the NNLO JR14 parametrization~\cite{Jimenez-Delgado:2014twa} for the $u_v - d_v$ that its prediction for the $\bar d(x) - \bar u(x)$ is clearly positive in all $x$, as seen in Fig.~\ref{fig:fig1}. In this way, if this sign-change occurs, we ensure that it is not resulted due to the selected PDFs. 
On the other hand, the CLAS Collaboration~\cite{Tkachenko:2014byy} has recently published the data for the neutron structure function $F_2^n$, and its ratio to the inclusive 
deuteron structure function $(F_2^n/F_2^d)$ as well as an updated extraction of Ref.~\cite{Baillie:2011za} for the ratio $R(x) = F_2^n/F_2^p$ from the BONuS 
experiment at Jefferson Lab. The data covers both the resonance and deep-inelastic regions including a wide range of $x$ for Q$^2$ between $0.7$ and $5$ GeV$^2$
and invariant mass $W$ between $1$ and $2.7$ GeV. In this way, the term $F^p_2(x) - F^n_2 (x)$  in Eq.~\eqref{eq2} can be calculated from
the data for the ratio $R(x)$ and by using the parametrization of $F_2^d(x)$ from Ref.~\cite{Arneodo:1995cq}, according to the 
following relation
\begin{equation}
F^p_2 - F^n_2 = 2F_2^d(1 - F_2^n/F_2^p)/(1 + F_2^n/F_2^p).
\label{eq3}
\end{equation}

\begin{figure}[t!]
	\centering
	\vspace{0.5 cm}
	\includegraphics[width=8.0cm]{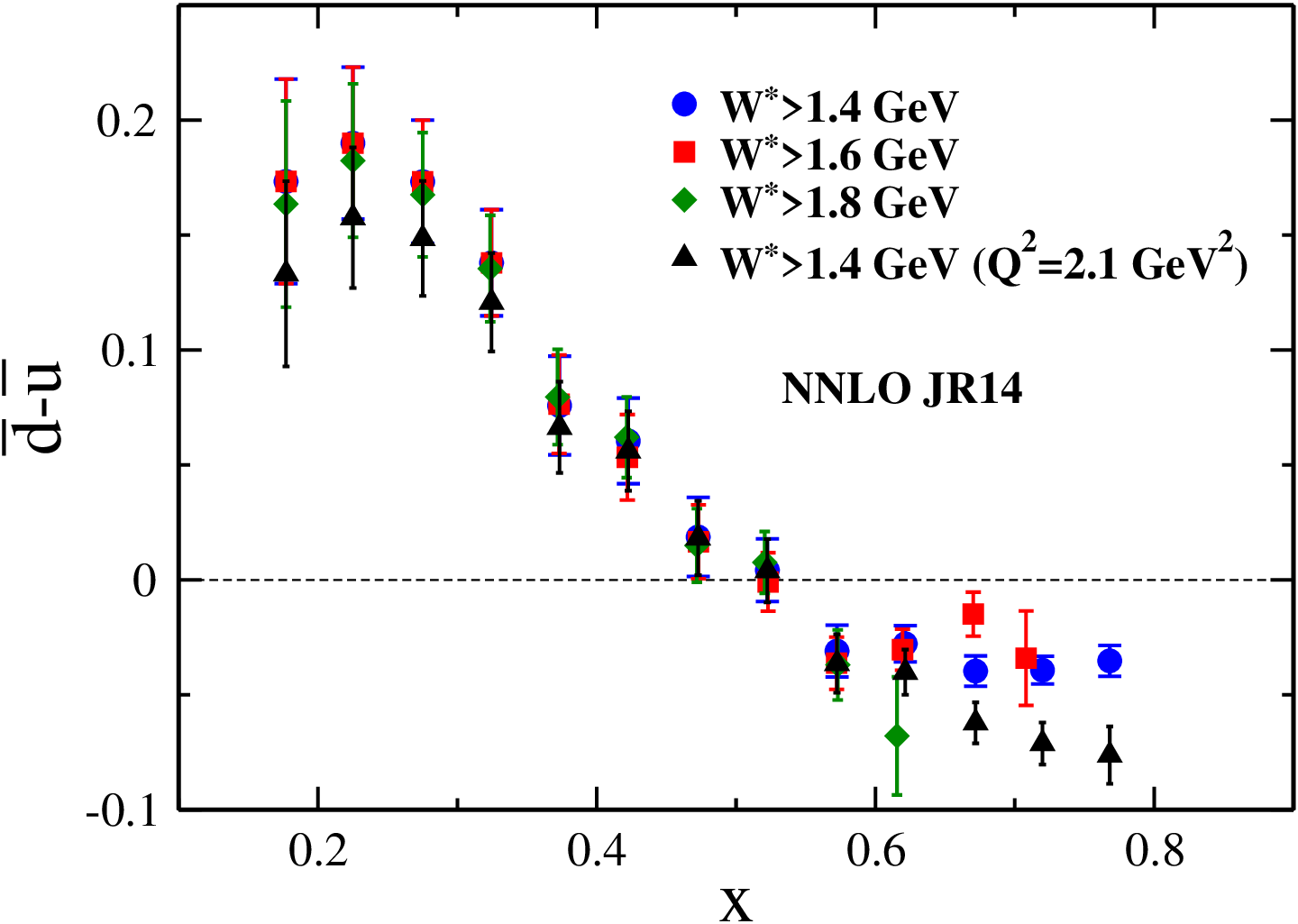}
	\caption{  The $\bar d(x) - \bar u(x)$ flavor asymmetry as a function of $x$. The results obtained by the NNLO JR14 parametrization~\cite{Jimenez-Delgado:2014twa} and the CLAS data~\cite{Tkachenko:2014byy} related to the three lower cuts on the range of final-state invariant mass $W^*$. The detailed explanation is given in the text. }
	\label{fig:fig2}
\end{figure}

\begin{figure}[t!]
	\centering
	\vspace{0.5 cm}
	\includegraphics[width=8.0cm]{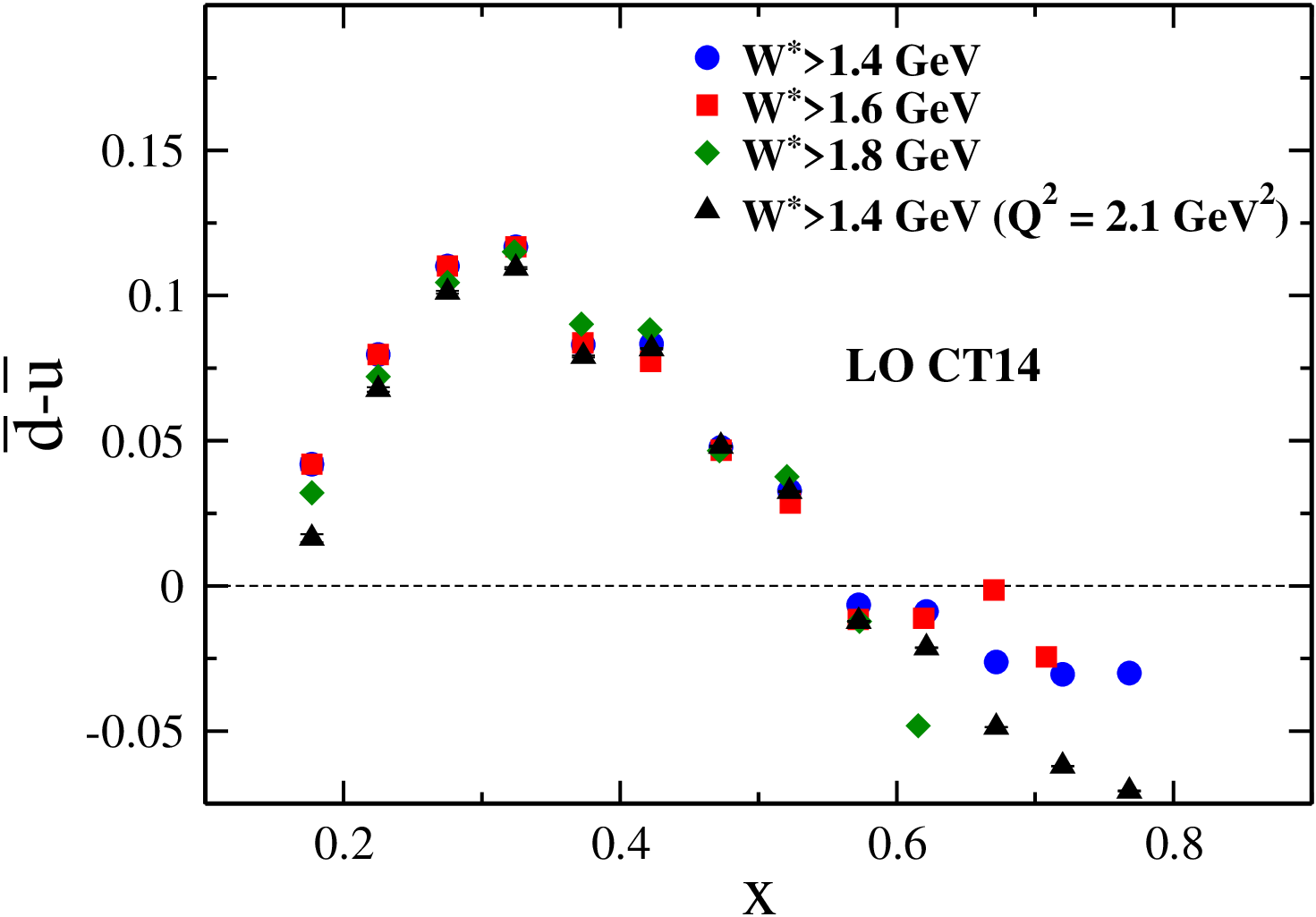}
	\caption{ As in Fig.~\ref{fig:fig2} but obtained from the LO CT14 parametrization~\cite{Dulat:2015mca}
		using the CLAS data~\cite{Tkachenko:2014byy}. The plot is related to the three lower cuts on the range of final-state invariant mass $W^*$.  }
	\label{fig:dbarmubarCLAS}
\end{figure}

Fig.~\ref{fig:fig2} shows our final results for the $\bar d(x) - \bar u(x)$ distribution, related to three lower cuts on the range of final-state invariant mass; $W^*>1.4$ GeV (blue circles), $W^*>1.6$ GeV (red squares) and $W^*>1.8 $ GeV (green diamonds). Note that, since the CLAS data are also $Q^2$-dependent and not related to a fixed value of Q$^2$, we have allowed all quantities
in Eqs.~\eqref{eq2} and \eqref{eq3} to be also $Q^2$-dependent. Therefore, the extracted $\bar d (x)- \bar u(x)$ data points in $x$ are related to the different $Q^2$ values approximately between $1$ and $4.5$ GeV$^2$. For example, for the case in which $W^*>1.6 $, the first and last data points are related to $Q^2=1.086$ and $4.259$ GeV$^2$, respectively.
However, we could also choose an average value for all data, i.e. $Q^2=2.1$ GeV$^2$. We examined this simplification and found it leads to an overall reduction 
in the magnitude of $\bar d(x)- \bar u(x)$, specifically, at small and large values of $x$. The related results have been shown in Fig.~\ref{fig:fig2}
as black triangles. To estimate the uncertainties, we have included both the uncertainties of $F_2^n/F_2^p$ and $F_2^d$
in our calculation for the $F^p_2 - F^n_2$ \eqref{eq3}, and also the JR14 PDFs uncertainties in the extraction of $\bar d (x)- \bar u(x)$ by using Eq.~\eqref{eq2}. As can be seen from Fig.~\ref{fig:fig2}, the high-quality data from the BONuS experiment leads to  rather smaller uncertainties.
It should be noted that Eq.~(\ref{eq2}) is extracted at the LO approximation but in our analysis, shown in Fig.~\ref{fig:fig2}, we used the NNLO PDF parametrization for more accuracy.
However, as we show in Fig.~\ref{fig:dbarmubarCLAS}, if one uses the LO PDF parameterizations from CT14~\cite{Dulat:2015mca}, the results show a sign-change as well.  

\begin{figure}[t!]
	\centering
	\vspace{0.5 cm}
	\includegraphics[width=8.0cm]{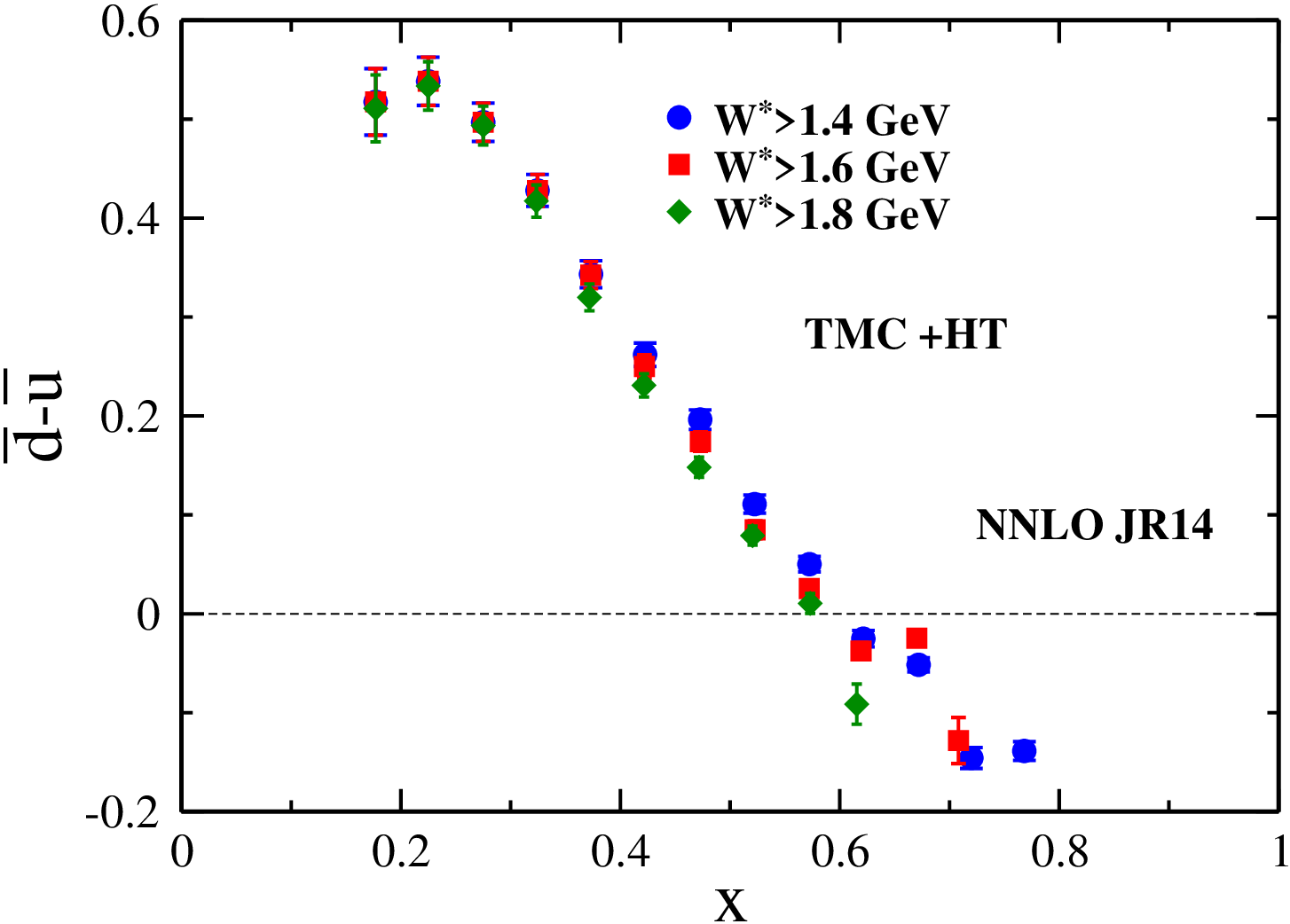}
	\caption{The $\bar d(x) - \bar u(x)$ asymmetry considering the TMC and  HT corrections. }
	\label{fig:dbarmubarCLAScorecnum}
\end{figure}

The last important issue which should be considered in our analysis is the effect of 
the nonperturbative target mass corrections (TMCs) and higher-twist (HT) terms.
At the region of low Q$^2$, nucleon mass correction cannot be neglected. Therefore, the power-suppressed corrections to the
structure functions can make an important contribution in some kinematical regions.
In addition to the pure kinematical origin TMCs, the structure functions also receive remarkable contributions from HT terms.
In the range of large values of $x$, their contributions are increasingly important.
In this respect, we examine the stability and reliability of our obtained results by including the TMCs as well as the HT terms which are particularly important
at the large-$x$ region and low Q$^2$.
Actually, since the CLAS measurements belong to the kinematical regions of $W \approx 2.7$ GeV and $Q^2 \approx 1 - 5$ GeV$^2$, and the Eq.~\eqref{eq1} might be too naive to use for the data points at such low $W$ and $Q^2$ regions, we should check the validity of our results by considering both the TMCs and HT term. In this regards, we follow the formalization presented in Refs.~\cite{Schienbein:2007gr} and \cite{Accardi:2009br} in order to taking into account the TMC and HT corrections in the structure functions of Eq.~\eqref{eq1}.
It should be also noted that for calculating the HT effect, we use the results presented in Table 3 of Ref.~\cite{Martin:2003sk}. Our final results have been shown in Fig.~\ref{fig:dbarmubarCLAScorecnum}, again for three lower cut values on $W^*$.  Comparing Figs.~\ref{fig:fig2} and \ref{fig:dbarmubarCLAScorecnum}, one can conclude that the TMCs and HT effect overall cause the results have larger value than before for positive area and the data points which were in the negative area have become more negative. Although, the TMCs and HT effect have paused some negative points to the positive area, we still have some data points which undergo the sign change. As a last point, note that if one uses the results obtained in Ref.~\cite{Botje:1999dj} for calculating the HT term, similar results will be achieved.

The most important conclusion of our analysis in this section is to show that the sign-change of $\bar d(x) - \bar u(x)$ occurs at large-$x$, as suggested by Peng \textit{et al.}~\cite{Peng:2014uea} in their analysis of the NMC DIS data for the $F^p_2 - F^n_2$~\cite{Amaudruz:1991at} and $F_2^d/F_2^p$~\cite{Arneodo:1996qe}, and also seen by the Drell-Yan experimental data measured at the Fermilab Experiment (E866)~\cite{Hawker:1998ty}. Although, this sign-change has occurred at $x \sim 0.5$, that is larger in comparison to the case of Drell-Yan data $ x\sim 0.3 $ (as shown in Fig.~\ref{fig:fig1}), but it seems reasonable because the CLAS data include very smaller values of $Q^2$ in comparison to the E866 data.
As another considerable point, note that in the definition of Eq.~\eqref{eq3} the nuclear effects in the deuteron, 
defined as $R_{\rm EMC}^d = F_2^d/(F_2^p + F_2^n)$, 
have been ignored. Actually, the nuclear corrections in the deuteron structure function are small
and usually are neglected in calculations. This fact is checked in the recent studies of the EMC effect in the deuteron by Griffioen \textit{et al.}~\cite{Griffioen:2015hxa}
through analyzing the recently published CLAS data at Jefferson Lab~\cite{Tkachenko:2014byy}. However, we recalculated the $\bar d(x) - \bar u(x)$
considering the nuclear corrections in the deuteron but only for the last data point that its related $R_{\rm EMC}^d$ ($=1.07$) is comparatively large, see Ref.~\cite{Griffioen:2015hxa}.
We found that it changes the result by 10\% so that the negativity of data at large-$x$ is still remaining.

%
\section{$\bar d(x) - \bar u(x)$ from BHPS model}\label{sec:three}

In this section, we present the results of our study for the $\bar d(x) - \bar u(x)$ in the basis of the BHPS model.
As was already mentioned in the Introduction, since the Gottfried sum rule has been violated by the NMC measurement~\cite{Amaudruz:1991at}, many theoretical studies based on the various models have yet been extended to explain the $\bar d(x) - \bar u(x)$ flavor asymmetry. Similar efforts have been also done in the case of
strange-antistrange asymmetry of the nucleon sea (for instance see Refs.~\cite{Cao:2003ny,Salajegheh:2015xoa,Vega:2015hti}). 
In recent years, Chang and Pang~\cite{Chang:2011vx} have demonstrated that a good description of Fermilab E866 data for the $\bar d(x) - \bar u(x)$ can be also achieved using the 
BHPS model~\cite{Brodsky:1980pb} for the intrinsic quark distributions in the nucleons. In the past three decades, the intrinsic quarks have been a subject of interest in
many researches including both intrinsic light and heavy quark components (see Refs.~\cite{Salajegheh:2015xoa,Brodsky:2015fna} and references therein). According
to the BHPS model that is pictured in the light-cone framework, the existence of the  five-quark Fock states $\vert uudq \bar{q} \rangle$ in the proton wave function is natural and 
the momentum distributions of the constituent quarks are given by 

\begin{equation}
	P(x_1, \cdots ,x_5) = N \frac{\delta \left(1-\sum\limits_{i=1}^5 x_i\right)}{\left(m_p^2- \sum\limits_{i=1}^5 \frac{m_i^2}{x_i}\right)^2},
	\label{eq4}
\end{equation}
where $m_p$ and $m_i$ refer to the masses of the proton and quark $i$, and $x_i$ stands for the momentum fraction carried by quark $i$. It should be noted that in Eq.~\eqref{eq4}
the effect of the transverse momentum in the five-quark transition amplitudes is neglected and the normalization factor $N$  is also determined through the following condition
\begin{equation}
	\int dx_1 \cdots dx_5 P(x_1, \cdots ,x_5)\equiv \mathcal{P}^{q\bar{q}}_5,
	\label{eq5}
\end{equation}
where $\mathcal{P}^{q \bar{q}}_5$ is a probability to find the $\vert uud q\bar{q}\rangle$-Fock state in the proton. Considering Eq.~\eqref{eq4}, one can integrate over $x_1, x_2, x_3$ and $x_4$ to obtain the 
$ \bar q $-distribution in the proton. As was mentioned in Ref.~\cite{Brodsky:1980pb}, the probability of the five-quark Fock state is proportional to $1/m_q^2$, where $m_q$ is the mass of
$q(\bar{q})$ in the Fock state $\vert uudq \bar{q} \rangle$. Although, the BHPS model prediction for the $\mathcal{P}^{q \bar{q}}_5$ is suitable when the quarks are heavy, we expect that the light five-quark states have a larger probability in comparison to the heavy five-quark states.

It is worth noting that the BHPS model was applied, at first, for calculating the intrinsic charm distribution \cite{Brodsky:1980pb}. However, 
Chang and Pang~\cite{Chang:2011vx} generalized it to the light five-quark states to calculate their intrinsic distributions in the proton and also to extract their probabilities ($\mathcal{P}^{q \bar{q}}_5$) using available experimental data. It is interesting to note that they obtained different values for the $\mathcal{P}^{d \bar{d}}_5$ and $\mathcal{P}^{u \bar{u}}_5$ and therefore they extracted $\bar d(x) - \bar u(x)$ distribution. This may leads us to a new idea so that we can chose different masses for down and up quarks in the BHPS formalism.
To make this point more clear, note that in one hand the $\mathcal{P}^{q\bar{q}}_5$ is proportional to $1/m_q^2$ and on the other hand, Eq.~\eqref{eq4} completely depends on the constituent quark masses, so these facts inevitably lead to the difference masses for the up and down quarks. Moreover, from~\cite{Chang:2011vx}, since $\mathcal{P}^{d \bar{d}}_5 (=0.294) $ is larger than $\mathcal{P}^{u \bar{u}}_5 (=0.176)$  then one can conclude that $m_{d, \bar{d}}$ should be smaller than $m_{u,\bar{u}}$. Considering this assumption, if one evolve the $\bar d(x) - \bar u(x)$ distributions to the experimental data scale~\cite{Hawker:1998ty}, it will provide a sing-change at large value of $x$, $x>0.3$.

To prove our claim, we should determine the real masses of down and up sea quarks by fitting the available experimental data for the $\bar d(x) - \bar u(x)$. To this end,
considering the definition of the $\chi^2$-function as~\cite{Martin:2009iq}
\begin{equation}\label{maral}
	\chi^2 = \sum _i \frac{(\Delta_i^{\rm data} - \Delta_i^{\rm theory})^2}{(\sigma_i^{\rm data})^2} \,,
\end{equation}
we must minimize it to obtain the optimum values for the up and down quark masses. Here, $\Delta_i^{\rm data}$ is the experimental data for the $\bar d(x) - \bar u(x)$.  In our analysis we use the HERMES~\cite{Ackerstaff:1998sr} and E866~\cite{Hawker:1998ty} data which are the only available data for this quantity. In (\ref{maral}), the theoretical result for the $\bar d(x) - \bar u(x)$ distribution ($\Delta_i^{\rm theory}$) is obtained from the BHPS model
and $\sigma_i^{\rm data}$ is the experimental error related to the systematical and statistical errors as: $(\sigma_i^{\rm data})^2 = (\sigma_i^{\rm stat})^2 + (\sigma_i^{\rm syst})^2$.

In our calculation of  the theoretical result $\Delta_i^{\rm theory}$, the required probabilities of $\vert uud u\bar{u}\rangle $
and $\vert uud d\bar{d} \rangle $ states (in the proton) are taken from the recent analysis of Chang and Pang~\cite{Chang:2014lea} who have done their analysis by considering   the new measurements of HERMES Collaboration \cite{Airapetian:2013zaw} for the $ x(s+\bar s) $. The related values are $\mathcal{P}^{u\bar{u}}_5= 0.229$ and $\mathcal{P}^{d\bar{d}}_5= 0.347$
for $\mu = 0.3$ GeV and also $\mathcal{P}^{u\bar{u}}_5 = 0.178$ and $\mathcal{P}^{d \bar{d}}_5 = 0.296 $ for $ \mu = 0.5 $ GeV,
where $\mu$ is the initial scale for the evolution of the non-singlet $\bar d(x) - \bar u(x)$ distribution to the scale of experimental data.

In this analysis, we merely extract the value of $m_{d,\bar d}$ by performing a fit to the experimental data. In fact, it is not necessary to extract $m_{u, \bar u}$ 
from data analysis, because one can determine this quantity using the following equation 
\begin{equation}\label{eq7}
	m_{u,\bar u} = \frac{m_p - m_{d, \bar d}}{2}.
\end{equation} 
The equation above is obtained by the fact that the proton consists of two up quarks and one down quark in the ground state.\\
To minimize the $\chi^2$-function (\ref{maral}), we employ the CERN program {\tt MINUIT}~\cite{James:1975dr} and perform our analysis at the LO and next-to-leading order (NLO) approximations. For both LO and NLO, our results are evolved from the initial scales $\mu=0.3$ GeV and $\mu=0.5$ GeV to the experimental data scales (Q$^2 = 54$ GeV$^2$ for the E866 data and Q$^2=2.5$ GeV$^2$ for the HERMES data). In Table \ref{tab1}, our results for $ m_{d, \bar d}$ along with the corresponding $\chi^2/{\rm d.o.f}$ values are presented for four scenarios, depend on the order of perturbative QCD and the initial scale applied.
\begin{table}[h!]
	\caption{The optimum values for the $d$-quark mass along with the corresponding $\chi^2/{\rm d.o.f}$ values.}
	\centering
	\begin{tabular}{ l c c }		                                            	\hline
		Approach & $\chi^2/{\rm d.o.f}$   &  $m_{d,\bar d}$                    \\   \hline    \hline
		LO ($\mu=0.3$)  & 6.3145  & \qquad 0.2020 $\pm$ 7.3357 $\times10^{-5}$ \\
		NLO ($\mu=0.3)$ & 1.0682  & \qquad 0.2779 $\pm$ 4.7401 $\times10^{-3}$ \\
		LO ($\mu=0.5$)  & 11.2947 & \qquad 0.2020 $\pm$ 5.1204 $\times10^{-5}$ \\
		NLO ($\mu=0.5)$ & 4.4402  & \qquad 0.2020 $\pm$ 8.3806 $\times10^{-5}$ \\
		\hline
	\end{tabular}
	\label{tab1}
\end{table}

According to Table~\ref{tab1} and Eq.~(\ref{eq7}), the possible values for the $m_{d,\bar d}$  are smaller than the $m_{u,\bar u}$ in all scenarios applied. 
As it can be seen from Table~\ref{tab1}, the value of $\chi^2/{\rm d.o.f}$ for the NLO approach considering the initial scale $\mu=0.3$~GeV is better than the other approaches. 
Another interesting point, shown in Table~\ref{tab1}, is that the values obtained for the $m_{d,\bar d}$ are the same when different scenarios are applied, i.e. LO ($\mu=0.3$ GeV), LO ($\mu=0.5$ GeV) and NLO ($\mu=0.5$ GeV). Considering  Table~\ref{tab1} and Eq.~(\ref{eq7}), our expectation value of the up quark mass is  $m_{u, \bar u}=0.330$~GeV using the second scenario where $\mu=0.3$~GeV is considered at NLO and one has $m_{u,\bar u}=0.368$~GeV considering other three scenarios. 

\begin{figure}[t!]
	\centering
	\vspace{0.5 cm}
	\includegraphics[width=8cm]{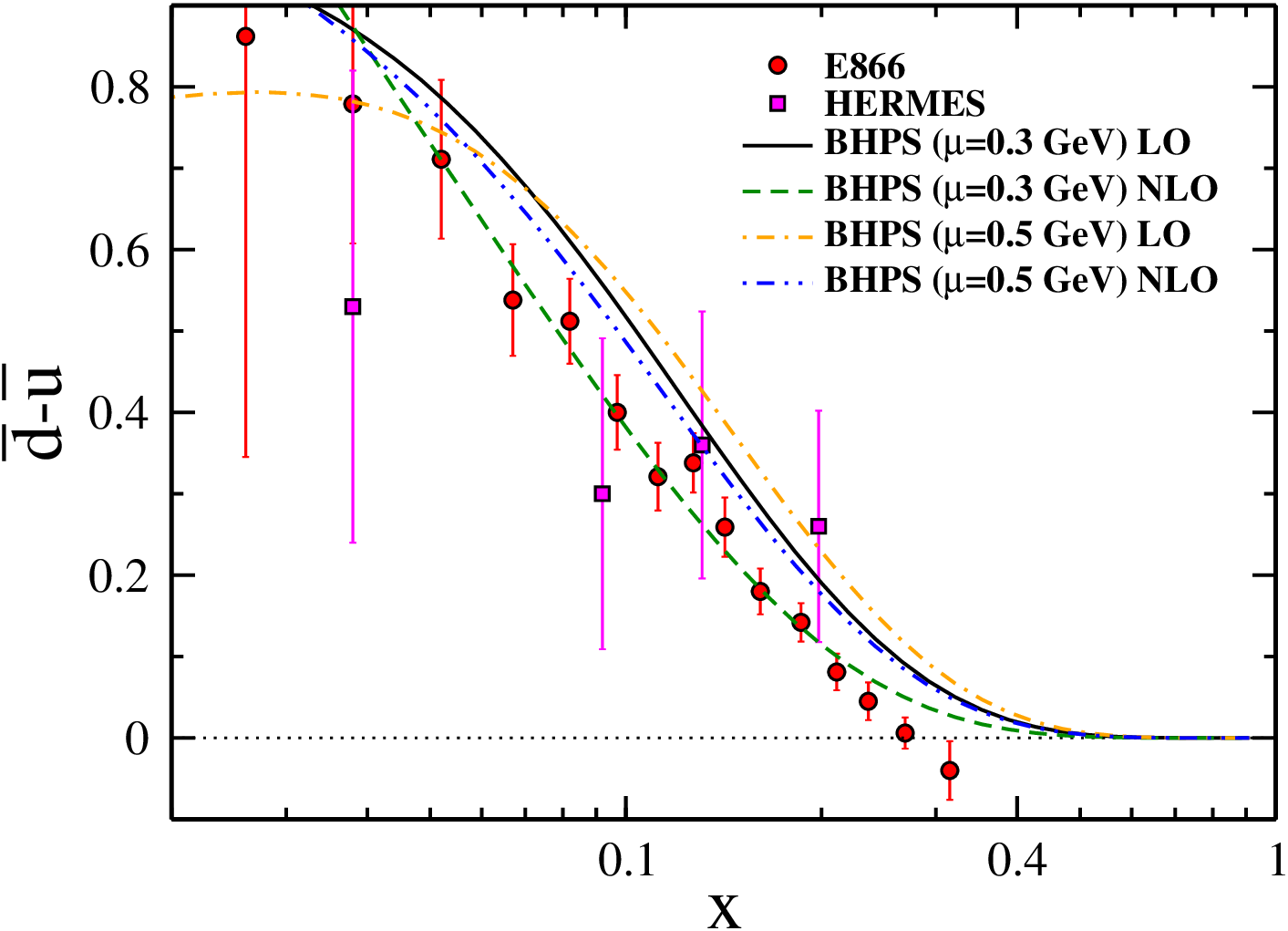}
	\caption{ A comparison between the experimental data from the HERMES~\cite{Ackerstaff:1998sr} and E866~\cite{Hawker:1998ty} collaborations 
		and the theoretical results obtained for $\bar d(x) - \bar u(x)$ in four situations, using the BHPS model with masses listed in Table~\ref{tab1}. }
	\label{fig:fig3}
\end{figure}

We have provided a code that gives the $\bar{d}$ and $\bar{u}$ intrinsic quark distributions in the proton for any arbitrary down quark mass and momentum fraction $x$ (see Appendix). Now, we can recalculate the BHPS model for the $\bar d(x) - \bar u(x)$ distribution using the new 
masses extracted for the up and down sea quarks. Because, the minimum value of $\chi^2/{\rm d.o.f}$ appears in the NLO scenario for $\mu=0.3$, we expect that this scenario leads to a more convenient consistency with the experimental data. Fig.~\ref{fig:fig3} shows a comparison between the experimental data and obtained results for the $\bar d(x) - \bar u(x)$ in four scenarios, using the BHPS model with the masses listed in Table~\ref{tab1}. Actually, these results show that our assumption is correct so choosing a smaller mass for the down quark is logical.

\begin{figure}[b!]
	\centering
	\vspace{0.5 cm}
	\includegraphics[width=8cm]{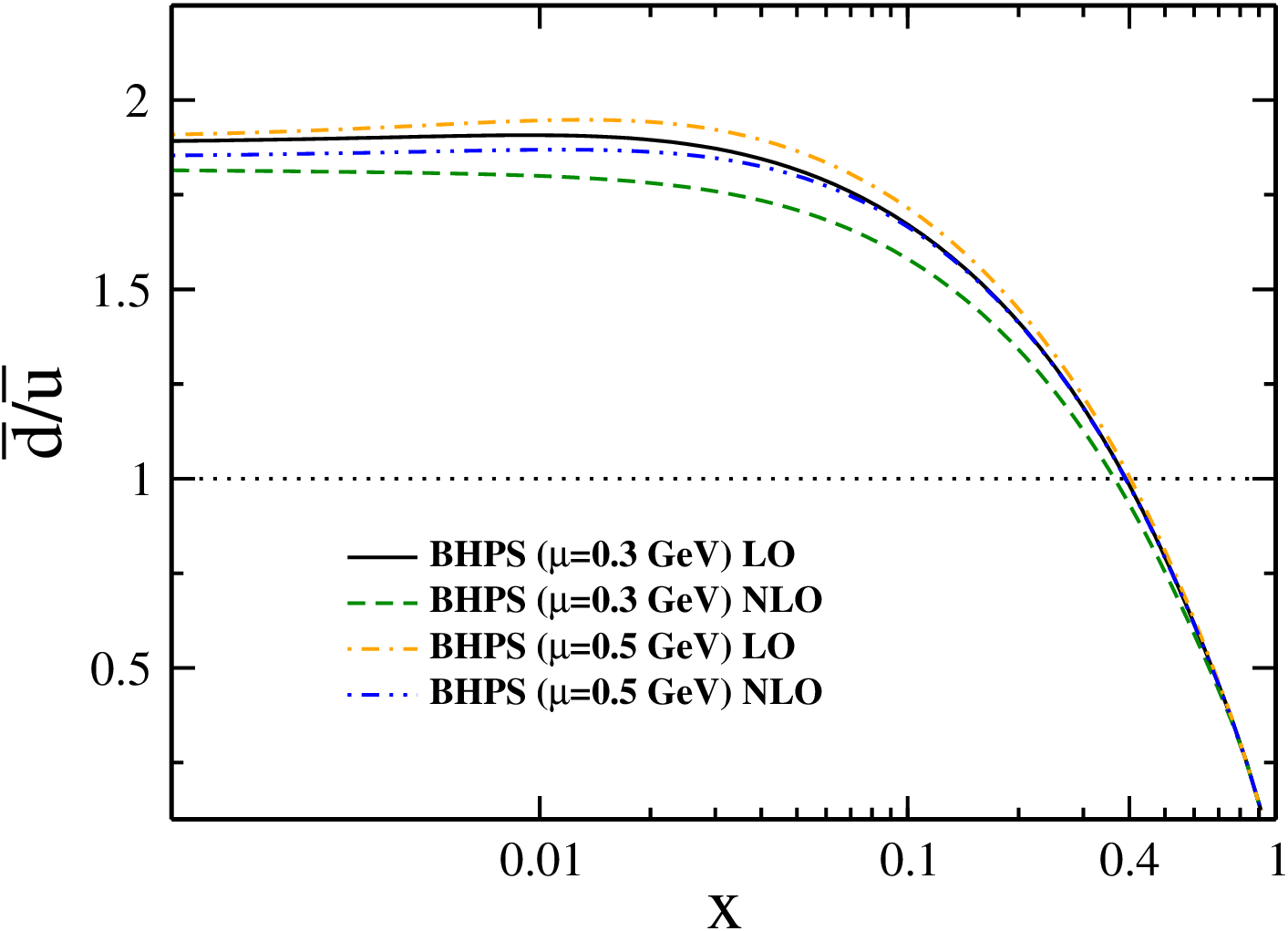}
	\caption{ $\bar{d}(x)/\bar{u}(x)$ versus $x$ which obtained in four situations, using the BHPS model with masses listed in Table~\ref{tab1}. }
	\label{fig:Plotdoveru}
\end{figure}
Another interesting finding has been achieved from our analysis is that the evolved distributions have a singe-change at the large value of $x$. The observed difference between $\bar d(x)$ and $\bar u(x)$ in this study for large value of $x$ is not significant as presented in Fig.~\ref{fig:fig3}. In this regard, for showing this sign-change, we have plotted the $\bar{d}(x)/\bar{u}(x)$ distribution as a function of $x$ for four analyzed scenarios. According to Fig.~\ref{fig:Plotdoveru}, at $x \gtrsim 0.33$ and for all approaches, the ratio of $\bar{d}(x)/\bar{u}(x)$ is smaller than 1.
From Fig.~\ref{fig:Plotdoveru} one can conclude that, for the NLO scenario and for $\mu = 0.3$ GeV, the corresponding curve fall down faster than the others.
The sign-change presented in this study have a number of important implications for future practice, and hence, any possible future studies on $\bar d(x) - \bar u(x)$ using the new and up-to-date experimental set up are most welcome. 

%
\section{Summary and Conclusion}\label{sec:four}

The experimental data taken from a Drell-Yan experiment by FNAL E866/NuSea collaboration~\cite{Hawker:1998ty} can be recognized as a cleanest
evidence for the violation of the Gottfried sum rule and the existence of the $\bar d(x) - \bar u(x)$ flavor asymmetry in the nucleon sea. Furthermore,
these data suggest a sign-change for the $\bar d(x) - \bar u(x)$ at $x \sim 0.3$. Recently, by analyzing the DIS data, Peng \textit{et al.}~\cite{Peng:2014uea} has presented an independent evidence for the $\bar d(x) - \bar u(x)$ sign-change at $x \sim 0.3$. They have showed that in addition to the Drell-Yan data,
the analysis of the NMC DIS data for the $F^p_2 - F^n_2$~\cite{Amaudruz:1991at} and $F_2^d/F_2^p$~\cite{Arneodo:1996qe} can also lead to a negative value for the $\bar d(x) - \bar u(x)$ at $x \gtrsim 0.3$.
They have also discussed the significance of this sign-change and the fact that none of the current theoretical models
can predict this effect. 
Following their studies, we have investigated this behavior in the DIS data analysis from other experiments.
Then, we have tried to found the $x$-position of the $\bar d(x) - \bar u(x)$ in which the sign-change occurs. In the following, we estimated 
the magnitude of the  negative area of the $\bar d(x) - \bar u(x)$ distribution. We have also enriched our formalism by considering the nonperturbative TMCs and HT terms. As a result, we fount that using the updated CLAS collaboration data for the structure function ratio $F_2^n/F_2^p$~\cite{Tkachenko:2014byy} 
the extracted $\bar d(x) - \bar u(x)$ undergoes a sing-change and becomes negative at large values of $x$, as suggested by Drell-Yan E866 data.

Then, we have used in the following the BHPS model~\cite{Brodsky:1980pb} to calculate the $\bar d(x) - \bar u(x)$ distribution. According to the BHPS prediction, we assumed that the probability of the  Fock state $\vert uudq\bar{q}\rangle$ in the proton wave function is proportional to $1/m_q^2$, where $m_q$ is the mass of
$q(\bar{q})$ in five quark Fock state. Under this assumption, the $d (\bar d)$ quark has a smaller mass than the $u (\bar u)$ quark in the proton.
To prove that, we obtained the real masses for the down and up sea quarks by fitting the available experimental data. We considered the $\chi^2$-function and minimized it
to obtain the optimum down and up sea quarks masses. Our calculations have been done in four scenarios: leading- and next-to-leading order approximations considering  two different initial scales 
$\mu = 0.3$ GeV and $\mu = 0.5$ GeV. Our results obtained from data analysis confirm the accuracy and correctness of our assumption.

The following short conclusions can be drawn from the present study.
As a short summary, the present results are significant in at least two major respects. First, we have found that the $\bar d(x) - \bar u(x)$ distribution with the new extracted masses, is in good agreement with the available dn up-to-date experimental data.
In addition, unlike the previously performed theoretical studies~\cite{Kumano:1997cy,Garvey:2001yq,Peng:2014hta}, our results show a sign-change on the $\bar d(x) - \bar u(x)$ distribution. The latter one is the more significant finding emerge from this study.
Any further information both on theory and exprimental observables on $\bar d(x) - \bar u(x)$ asymmetry would help us to establish a greater degree of accuracy on this matter. 
Theses are important issues for future research, and hence, further studies with more focus on $\bar d(x) - \bar u(x)$ asymmetry are therefore suggested.

%
\section*{Acknowledgments}

Authors are thankful of School of Particles and Accelerators, Institute for Research in Fundamental Sciences (IPM) for financially support of this project. 
Hamzeh Khanpour also gratefully acknowledge the University of Science and Technology of Mazandaran for financial support provided for this research.

%
%
\appendix

\section*{Appendix: The $\bar{d}$ and $\bar{u}$ intrinsic distributions}
We have provided a {\tt FORTRAN} package containing the $\bar{d}$ and $\bar{u}$ intrinsic distributions using the BHPS model for any arbitrary down quark mass and momentum fraction $x$ which can be obtained via e-mail from the authors. Note that, in this code the probabilities $\mathcal{P}^{d \bar{d}}_5$ and $\mathcal{P}^{u \bar{u}}_5$ have not been multiplied by distributions so one can choose any arbitrary probabilities. Furthermore, the up quark mass $m_{u, \bar u}$ is obtained from Eq.~(\ref{eq7}), automatically.

%
%

\end{document}